\documentclass{article}

\usepackage{graphics}

\newcommand{\xt}{{x}}

\newcommand{\kt}{\bar{k}}

\newcommand{\et}{\bar{\epsilon}}

\newcommand{\ap}{\alpha}

\newcommand{\bt}{\beta}

\newcommand{\gm}{\gamma}

\newcommand{\lm}{\lambda}

\newcommand{\ep}{\epsilon}

\newcommand{\as}{\alpha_{\rm s}}

\newcommand{\Nc}{N_{\rm c}}

\newcommand{\CF}{C_F}

\newcommand{\bq}{\bar{q}}

\newcommand{\be}{\begin{equation}}

\newcommand{\ee}{\end{equation}}

\newcommand{\ba}{\begin{eqnarray}}

\newcommand{\ea}{\end{eqnarray}}

\newcommand{\dxt}[1]{\frac {d^2 \xt _{#1}} {2 \pi}}

\begin{document}

\thispagestyle{empty}

\begin{flushleft}
 ~
\end{flushleft}

\vspace{3cm}

\centerline{{\Large \bf {
Problems with proton in the QCD dipole picture
}}}

\vspace{1cm}

\centerline{M.~Prasza{\l}owicz\footnote{Present address: Institute
of Theor. Phys. II, Ruhr-University Bochum, D-44780 Bochum, Germany}
 and A.~Rostworowski}

\smallskip
\begin{center} {\sl { 
Particle Theory Department, \\
Institute of Physics, \\
Jagellonian University, \\
Reymonta 4, \\
30-049 Krak{\'o}w,  Poland
}}
\end{center}
\bigskip
\vfil
\abstract{
The soft gluon part of a proton wave function is investigated and
compared with an {\em onium} case. It is argued that at every step of
the gloun cascade new color structures appear. 
Dipole equation kernel emerges when a diquark limit is assumed.
}
\vfill

\newpage

\section{Introduction}

A soft gluon evolution in QCD is governed by a so called BFKL
equation which has been originally derived in \cite{BFKL1, BFKL2,
BFKL3}.  In a series of papers \cite{Nikolaev1} by Nikolaev and
Zaharkov and independently \cite{Mueller1, Mueller2, Mueller3} by
Mueller and collaborators a dipole picture of the gluon cascade has
been developed and shown to be equivalent to the BFKL evolution. For
its simplicity and probabilistic interpretation the QCD dipole
picture  has been successfully applied to describe proton structure
function at low $x$ measured at HERA \cite{Peschanski1, Peschanski2},
to proton-nucleon scattering \cite{BWC1} to photon-photon scattering
\cite{BWC2} and Pomeron phenomenology \cite{BP,Bialas}.

The most convenient way to introduce the dipole picture is to
consider an $onium$:  a $q \bq$ state of two heavy fermions. The
$onium$ wave function consists of the original fermions and many
gluons emitted during the time evolution.  In order to calculate the
$n$ gluon component of the wave function one employs:  

1. a light-cone perturbation theory and leading logarithmic approximation
for the longitudinal phase-space integrations;  

2. a large $\Nc$
expansion in which every gluon line in color space can be represented
as a $q \bq$ pair. 

In the first step (1 gluon component of the
wave function) a quark component of the gluon and the heavy antiquark
from the original $onium$ which form a color singlet are said to
constitute a dipole. A second dipole is formed from an antiquark
component of the gluon and the heavy quark from the $onium$.  The key
observation leading to the formulation of the dipole picture is that
in the next steps (2 and more gluon component of the wave function)
each dipole emits subsequent gluons independently of other dipoles;
in other words there is no interference between the different dipoles.
This leads to a factorization of a kernel describing gluon emission
from a dipole.

In this way real gluon emissions are taken into account. Total
probability of real emissions is UV divergent but IR finite. Virtual
corrections, which have to be included as well, make this probability
finite and the complete kernel is identical to the one of BFKL.

In \cite{Peschanski1, Peschanski2} it was assumed that $q \bq$
(dipole) configuration can be found in a proton and this dipole
configuration gives the dominant contribution to the soft gluon
density in a proton.  This assumption leads to the same soft gluon
density in the case of a proton as in the case of an {\em onium} 
(up to the
normalization factor).  Thus it becomes tempting to get this result
by explicit calculation.  Indeed, one can naively expect that a 3 (or
rather $\Nc$) quark configuration in a color singlet state surrounds
itself by dipoles during the evolution of the gluon cascade in close
resemblance to the {\em onium} case.

In this letter we show that this naive picture breaks down.  The
reason is quite simple: there are interference diagrams between 
dipoles and a proton which  cannot be neglected as it was in the case
of the {\em onium}. There the interference diagrams between different
dipoles could have been discarded since they were non-leading in
$\Nc$. However, in the case of a proton a number of quarks is $\Nc$ and
this factor enhances the non-leading interference contributions. As a
result there is no universal emission kernel which factorizes in
subsequent gluon emissions and moreover the complicated color
structures emerge.  Whether a simple probabilistic picture can be
recovered by introducing higher color multipoles remains still an
open question.

It is however instructive to investigate a diquark limit, $i.e.$ a
limit in which 2 (or $\Nc - 1$) quarks are localized in one point.
In that case the original proton reduces to a quark-diquark system
which is in fact identical to an {\em onium} state. We have checked 
by explicit calculation that this diquark configuration of  valence 
quarks in a proton leads to the soft gluon density of a dipole type.

In Section 2, we calculate the wave function of a proton in analogy
with \cite{Mueller1}. Then the square of a proton wave function is
compared with the {\em onium} case.  In Section 3, we show that in the
limit where two valence quarks are localized in the same transverse
position (diquark approximation) the dipole equation emerges. 
In Section 4 we shortly summarize our findings.

\section{Wave function of a proton}

In analogy with Refs.\cite{Mueller1, Mueller3} we decompose the proton 
state $| \Psi >$ in the basis of the eigenstates of the QCD evolution
hamiltonian $H_0$. The $3+n$ particle (three quarks and $n$ soft gluons)
component of $| \Psi >$ is defined as: 

\ba
\label{protonPsi}
{\psi^{(n)}}^{~ijk~{a_1 \dots a_n}}_{{\ap \bt \gm}~{\lm_1 \dots \lm_n}}
(k_0, \dots k_{2+n})~~~~~~~~~~~~~~~~~~~~~~~~~~~~~~~~~~~~~~~~~~~~~~~~
\nonumber \\ = 
< k_0,\ap, i; k_1, \bt, j; k_2, \gm, k; 
k_3,\lm_1, a_1; \dots k_{2+n}, \lm_n, a_n | \Psi >.  
\ea
$k_0, ~k_1$ and $k_{2}$ denote quark momenta, whereas
$k_{3}, \dots k_{2+n}$ correspond to gluons. Indices
$\ap, \bt, \gm$ and $ \lm_1, \dots \lm_n$ denote quark and gluon
polarizations respectively, and 
$i, j, k, a_1, \dots a_n$ -- colors of three valence quarks and $n$
gluons.

We take three valence quarks with no soft gluons to be the
initial proton state (fig.1). Black dot on fig.1 denotes the
antisymmetrization of the valence quarks in color space. It is convenient to 
factor out the antisymmetric tensor in $\psi^{(0)}$ explicitly:
\be
{\psi^{(0)}}^{~ijk}(k_0, k_1, k_2) = 
\frac{1}{\sqrt{3!}}\, \ep_{ijk}{\psi^{(0)}} (k_0, k_1, k_2).
\ee
In what follows we shall also suppress spinor indices ${\ap \bt \gm}$.

\begin{figure}
\label{fig1}
\begin{center}
\includegraphics{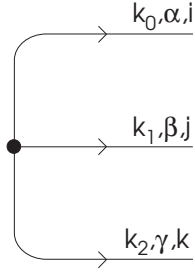}
\caption{Initial proton state - three valence quarks with no soft gluons}
\end{center}
\end{figure}

Starting with three valence quarks, described by 
$\ep_{ijk} \psi^{(0)} (k_0, k_1, k_2)$, we can calculate 1-soft gluon
component of a proton wave function. Summing contributions of the
three graphs in fig.2 we find (in the first order in 
strong coupling constant $g$):

\ba
{\psi^{(1)}}^{~ijk~a_1}_{~~~~~\lm_1}(k_0, \dots k_3) 
                    = i 2 \sqrt{2} g \frac {\kt_3 \cdot \et_{\lm_1}} {\kt_3^2} 
\left[ 
  \frac {\epsilon_{i'jk} (T^{a_1})^i_{i'}} {\sqrt{3!}} 
                                      \psi^{(0)} (k_0 + k_3, k_1, k_2) \right.
\nonumber\\ \left. 
+ \frac {\epsilon_{ij'k} (T^{a_1})^j_{j'}} {\sqrt{3!}} 
                                       \psi^{(0)}(k_0, k_1 + k_3, k_2) 
+ \frac {\epsilon_{ijk'} (T^{a_1})^k_{k'}} {\sqrt{3!}} 
                                       \psi^{(0)}(k_0, k_1, k_2 + k_3) \right].
\ea
Here we have used a light-cone decomposition of the momenta, with $\kt_i$ 
being the transverse component of $k_i$. It is convenient to work in a mixed 
momentum-space representation performing Fourier transform in $\kt_i$. 
The corresponding 2-dimensional
transverse position vectors are subsequently denoted by 
$x_i$, whereas $z_i$ correspond to the fractions of the ``+'' components 
of momenta with respect to the ``+'' component of the initial proton momentum.
The details of the kinematics can be found in Ref.\cite{Mueller1}. 
The 1-gluon component of the proton wave 
function takes the following form in the mixed momentum-space representation:
\ba
{\psi^{(1)}}^{~ijk~a_1}_{~~~~~\lm_1}(\xt_0, \dots \xt_3; z_0, \dots z_3) 
 = - \frac{\sqrt{2}}{\pi} g \psi^{(0)}(\xt_0, \xt_1, \xt_2; z_0, z_1, z_2)
\nonumber\\
\left[
 \frac{\epsilon_{i'jk} (T^{a_1})^i_{i'}}{\sqrt{3!}} 
 \frac{\xt_{30}\cdot\et_{\lm_1}} {\xt_{30}^2} 
+\frac{\epsilon_{ij'k} (T^{a_1})^j_{j'}}{\sqrt{3!}} 
 \frac{\xt_{31}\cdot\et_{\lm_1}}{\xt_{31}^2} 
+\frac{\epsilon_{ijk'} (T^{a_1})^k_{k'}}{\sqrt{3!}} 
 \frac{\xt_{32}\cdot\et_{\lm_1}} {\xt_{32}^2} \right],
\ea
where $x_{ij}=x_i-x_j$ and $\et_{\lm_i}$ is transverse part of gluon
polarization vector.

\begin{figure}
\label{fig2}
\begin{center}
\includegraphics{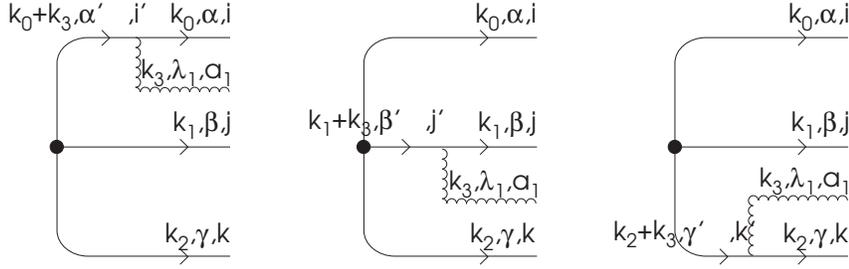}
\caption{One gluon component of a proton wave function}
\end{center}
\end{figure}

\begin{figure}
\label{fig3}
\begin{center}
\includegraphics{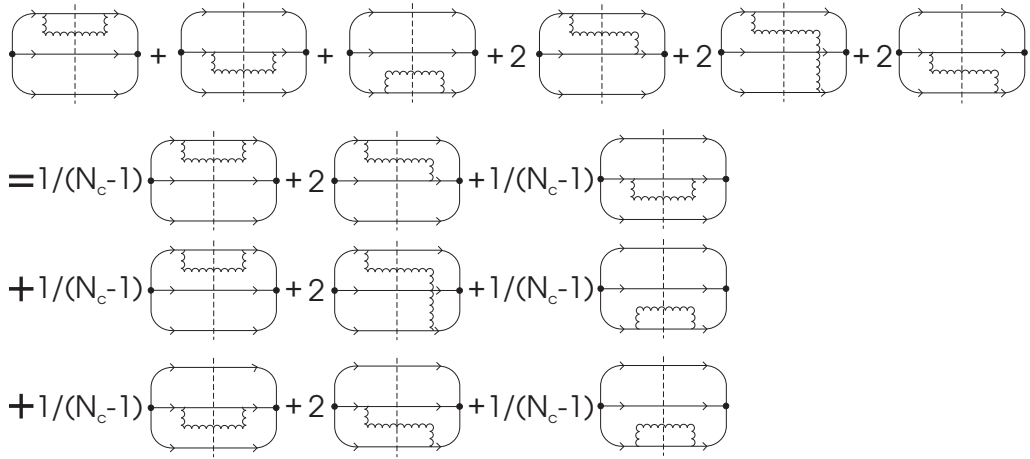}
\caption{Square of the one gluon component of a proton wave function for 
$\Nc=3$}
\end{center}
\end{figure}

Now we can calculate $\Phi^{(1)}$ -- square of the 1-gluon component 
of the proton wave function summed over gluon polarizations, which is 
depicted in the first row of fig.3. The color factor of the
noninterference graphs (gluon
emitted and absorbed by the same quark line -- first three graphs
of the first row of fig.3) equals $\CF$. The color factor of the
interference graphs (gluon exchanged between the different quark lines
-- last three graphs of the first row of fig.3) equals $\CF / (\Nc -
1)$. However we cannot neglect the interference diagrams, even in the
large $\Nc$ limit. That is because in the large $\Nc$ limit a
proton would be composed not from three but from $\Nc$ valence
quarks. Therefore effectively each non-interference diagram  has to
be split into $(\Nc - 1)$ 'copies' in order to match the interference
graphs. This is pictorially illustrated in fig.3 for $\Nc = 3$.
Each row in the second part of fig.3 sums up to a compact expression
resembling a dipole production by an {\em onium}. Summing up the 
contribution from all 3 (or rather $\Nc (\Nc - 1) / 2$) rows we get:
\ba
\label{proton Phi1}
\Phi^{(1)} (\xt_0, \xt_1, \xt_2, \xt_3; z_0, z_1, z_2, z_3) & =  &
  \Phi ^{(0)} (\xt_0, \xt_1, \xt_2; z_0, z_1, z_2)  \nonumber \\
 & &
\frac {8 \as} {\pi} \frac {\Nc + 1} {2 \Nc} \,
\left[ \frac {x_{10}^2} {x_{30}^2 x_{31}^2} 
     + \frac {x_{20}^2} {x_{30}^2 x_{32}^2} 
     + \frac {x_{21}^2} {x_{31}^2 x_{32}^2} \right].
\ea
It is important to note that interference diagrams make
the integral:
\be
\label{proton Phi1 int}
\int \dxt{3} \Phi ^{(1)} (\xt_0, \dots \xt_3; z_0 , \dots z_3)
\ee
IR finite (like in the dipole case). The UV divergence appears because
the $virtual$ corrections have not been taken into account (see
\cite{Mueller1}) and the integral (\ref{proton Phi1 int}) should be
appropriately regularized.

We can compare this result with the case of an {\em onium}.
\be
\label{onium Phi1}
\Phi _{\rm onium} ^{(1)} (\xt_0, \xt_1, \xt_2; z_0, z_2, z_3) 
= \Phi _{\rm onium} ^{(0)} (\xt_0, \xt_1; z_0, z_1) 
\frac {8 \as \CF} {\pi} \frac {x_{10}^2} {x_{20}^2 x_{21}^2},
\ee
gives the universal probability of a gluon emission from a dipole. 
In the large $\Nc$ limit a gluon line can be represented as a 
quark-antiquark pair in  color space. The {\em onium} with one gluon 
has then a straightforward interpretation of two dipoles and the 
probability (\ref{onium Phi1}) can be interpreted as a probability of 
a dipole (or equivalently an {\em onium}) splitting into two dipoles.

In the large $\Nc$ limit there is no interference between different 
dipoles, so once a gluon is emitted from one of the already existing 
dipoles we can again identify the resulting color structure as two 
new dipoles. Therefore the probability  of emission of the second 
(or $n$-{th})  gluon is  given by the sum of the  
probabilities of a gluon emission by two (or $n$) already existing 
dipoles. Thus for $\Phi ^{(2)} _{\rm onium}$ we get an expression:
\ba
\label{onium Phi2}
\Phi _{\rm onium} ^{(2)} (\xt_0,\dots \xt_3; z_0,\dots z_3) & = & 
\Phi _{\rm onium} ^{(0)} (\xt_0, \xt_1; z_0, z_1) \nonumber \\
 & &
\left( \frac {8 \as \CF} {\pi} \right) ^2 \frac {x_{10}^2} {x_{20}^2 x_{21}^2}
\left( \frac {x_{20}^2} {x_{30}^2 x_{32}^2} 
     + \frac {x_{21}^2} {x_{31}^2 x_{32}^2} \right),
\ea
where the first factor ${x_{10}^2}/ {x_{20}^2 x_{21}^2}$ describes a process
in which the original {\em onium} splits into 2 dipoles and the second term
in brackets describes the splitting of these 2 dipoles into 2 new 
dipoles each. This pattern is universal at every order of $\as$.

Similarly $\Phi ^{(1)}$ of eq.(\ref{proton Phi1})
can be interpreted as a probability of
an emission of the first soft gluon from a proton. Representing the
first gluon as a quark-antiquark pair, one might try to interpret the 
resulting state as a proton made of two original quarks and a quark part
of the gluon and a dipole made of an antiquark part of a gluon and the 
remaining quark of the original proton (fig.4a). In analogy with the 
{\em onium} case one could naturally expect that in the next step 
one would get: 

 1. a proton emitting the next gluon described by the kernel of 
eq.(\ref{proton Phi1}) and

 2. a dipole emitting the next gluon described by the kernel of
eq.(\ref{onium Phi1}). \\
Unfortunately it is not so.

\begin{figure}
\label{fig4}
\begin{center}
\includegraphics{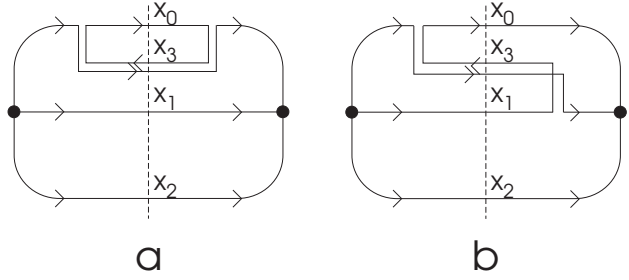}
\caption{Large $\Nc$ limit of the first gluon emission}
\end{center}
\end{figure}

We shall show this by explicit calculation of the square of the proton wave 
function. We shall do this {\em  exactly}, {\em i.e.} without escaping to 
the large $\Nc$ limit. The 2-gluon component of the proton wave 
function equals:
\ba
\label{proton psi2}
{\psi^{(2)}}^{~ijk~a_1 a_2}_{~~~~~\lm_1 \lm_2}(\xt_0, \dots \xt_4) =
- \frac {\sqrt{2}} {\pi} g  & & \left[ 
  (T^{a_2}) ^i _{i'} {\psi^{(1)}}^{~i'jk~a_1}_{\lm_1} (\xt_0, \dots \xt_3) 
  \frac {\xt_{40} \cdot \et_{\lm_2}} {\xt_{40}^2} \right.
\nonumber \\ 
& & 
+ (T^{a_2}) ^j _{j'} {\psi^{(1)}}^{~ij'k~a_1}_{\lm_1} (\xt_0, \dots \xt_3) 
  \frac {\xt_{41} \cdot \et_{\lm_2}} {\xt_{41}^2} 
\nonumber \\ 
& &
+ (T^{a_2}) ^k _{k'} {\psi^{(1)}}^{~ijk'~a_1}_{\lm_1}(\xt_0, \dots \xt_3) 
  \frac {\xt_{42} \cdot \et_{\lm_2}} {\xt_{42}^2} 
\nonumber \\ 
& & \left.
+ i f^{a_1 a_2 c} {\psi^{(1)}}^{~ijk~c}_{\lm_1}(\xt_0, \dots \xt_3) 
  \frac {\xt_{43} \cdot \et_{\lm_2}} {\xt_{43}^2} \right].
\ea

In order to calculate $\Phi ^{(2)}$ it is convenient to split the entire
expression into 2 parts. Contribution to $\Phi ^{(2)}$ coming from 
the second  gluon attached in all possible ways within a diagram of fig.4a 
equals:\footnote{Strictly speaking fig.4 represents the large $\Nc$ limit of
the first gluon emission, however, as al{\-}re{\-}ady said, we calculate 
2-gluon emission {\em exactly}.}
\ba
\label{Phi2 4a}
\Phi ^{(2)}_{4{\rm a}} (\xt_0, \ldots  \xt_4; z_0, \ldots  z_4) & & = 
\Phi ^{(0)} (\xt_0, \xt_1, \xt_2; z_0, z_1, z_2) 
\left( \frac {8 \as} {\pi} \right) ^2 {\CF}   \\
 & &\frac{1}{x_{30}^2}
\left[ 
  \frac {\Nc} {2} \frac {x_{30}^2} {x_{43}^2 x_{40}^2} 
 +\frac {\Nc + 1} {2 \Nc} 
\left( \frac {x_{31}^2} {x_{43}^2 x_{41}^2} 
     + \frac {x_{32}^2} {x_{43}^2 x_{42}^2} 
     + \frac {x_{12}^2} {x_{41}^2 x_{42}^2} \right) \right.
\nonumber \\ 
 & & \left. 
+ \frac {1} {2 \Nc (\Nc - 1)} 
\left( \frac {x_{31}^2} {x_{43}^2 x_{41}^2} 
     + \frac {x_{32}^2} {x_{43}^2 x_{42}^2} 
     - \frac {x_{01}^2} {x_{40}^2 x_{41}^2} 
     - \frac {x_{12}^2} {x_{41}^2 x_{42}^2} \right) \right] .
\nonumber \ea
In this case the probability of the second gluon emission
behaves as we would have a dipole ($x_3$, $x_0$), a proton ($x_3$,
$x_1$, $x_2$) and a sum of four (or $2(\Nc - 1)$) terms which in the
large $\Nc$ limit can be neglected. However the prefactor  
$1/{x_{30}^2}$ corresponds only to one part of the emission kernel
of the first gluon. Contribution to $\Phi ^{(2)}$  coming from the 
second gluon attached in all possible ways within a diagram 
of fig.4b equals:
\ba
\label{Phi2 4b}
\Phi ^{(2)}_{4{\rm b}} (\xt_0, \ldots  \xt_4; z_0, \ldots  z_4) & & =
\Phi ^{(0)} (\xt_0, \xt_1, \xt_2; z_0, z_1, z_2) 
\left( \frac {8 \as} {\pi} \right) ^2 \frac {\Nc + 1} {2 \Nc} 
\nonumber \\ & &
\left( - 2 \frac {\xt _{30} \cdot \xt _{31}} {x_{30}^2 x_{31}^2} \right) 
\left[ \frac{\Nc} {2} 
\left( \frac {x_{03}^2} {x_{40}^2 x_{43}^2} 
     + \frac {x_{13}^2} {x_{41}^2 x_{43}^2} 
     - \frac {x_{01}^2} {x_{40}^2 x_{41}^2} \right) \right.
\nonumber\\ 
 & & \left. 
     + \frac {\Nc + 1} {2 \Nc} 
\left( \frac {x_{01}^2} {x_{40}^2 x_{41}^2} 
     + \frac {x_{02}^2} {x_{40}^2 x_{42}^2} 
     + \frac {x_{12}^2} {x_{41}^2 x_{42}^2} \right) \right].
\ea
In this case we get a color structure which cannot be identified
with a proton and a dipole emitting  a gluon. The probability of the 
second gluon emission behaves as we would have the original proton 
($x_0$, $x_1$, $x_2$) and two dipoles ($x_3$, $x_0$ and $x_3$, $x_1$) with
subtraction of another dipole ($x_0$, $x_1$). This is a signature of a new
color structure. Here the prefactor   corresponding to the first emission
reads:
$- 2\; \xt _{30} \cdot \xt _{31} / {x_{30}^2 x_{31}^2} $. The fact that even 
in the large $\Nc$ limit expressions in square brackets in eq.(\ref{Phi2 4a})
and eq.(\ref{Phi2 4b}) are different, makes it impossible to collect the
prefactors into one compact expression like eq.(\ref{proton Phi1}).  

To get the full expression for $\Phi^{(2)}$ one has to sum all (see fig.3)
contributions from three (or $\Nc$) diagrams of the type
of fig.4a, where the first gluon is emitted and absorbed by the same quark
in the initial proton and from three (or $\Nc (\Nc - 1) / 2$) diagrams of 
the type of fig.4b, where the first gluon line  extends between two different 
quark lines of the original proton. On the top of the first gluon the
second gluon line should be added accordingly. Thus one gets:
\ba
\label{Phi2}
&&
\Phi ^{(2)} (\xt_0, \dots, \xt_4; z_0, \dots z_4) =
\nonumber \\
&&
\Phi ^{(2)}_{4{\rm a}} (\xt_0, \xt_1, \xt_2, \xt_3, \xt_4; z_0, \ldots  z_4) + 
\Phi ^{(2)}_{4{\rm a}} (\xt_1, \xt_0, \xt_2, \xt_3, \xt_4; z_0, \ldots  z_4) + 
\Phi ^{(2)}_{4{\rm a}} (\xt_2, \xt_1, \xt_0, \xt_3, \xt_4; z_0, \ldots  z_4) + 
\nonumber \\
&&
\Phi ^{(2)}_{4{\rm b}} (\xt_0, \xt_1, \xt_2, \xt_3, \xt_4; z_0, \ldots  z_4) +
\Phi ^{(2)}_{4{\rm b}} (\xt_0, \xt_2, \xt_1, \xt_3, \xt_4; z_0, \ldots  z_4) +
\Phi ^{(2)}_{4{\rm b}} (\xt_1, \xt_2, \xt_0, \xt_3, \xt_4; z_0, \ldots  z_4)
\nonumber \\
\ea
Although the full expression is IR finite there is no factorization of 
the first gluon emission as given by eq.(\ref{proton Phi1}). It seems 
to us that this might be due to the fact that at every step of the gluon 
cascade new color structures appear which cannot be reduced to one proton 
and a number of dipoles.

\section{Diquark limit and dipole equation}

It is not clear to us whether the assumption, made in 
Refs.\cite{Peschanski1, Peschanski2},
that up to a normalization
factor there is no difference between a proton and an {\em onium} soft 
gluon dynamics can be justified
in  general case. However, if we assume that two valence quarks are
localized in the same transverse position we recover the dipole equation.
This is is of course an expected feature, since two (or $\Nc-1$) quarks in
antisymmetric state behave like an antiquark.   

It is, however, instructive to check this on the example of $\Phi^{(2)}$.
Putting $\xt_1 = \xt_2$ in the expression (\ref{Phi2}) for $\Phi ^{(2)}$
and remembering about combinatory factors we get:
\ba
\Phi ^{(2)}(\xt_0, \xt_1 = \xt_2, \xt_3, \xt_4; z_0, \dots z_4) 
 & &  =   \Phi^{(0)}(\xt_0, \xt_1 = \xt_2; z_0, z_1, z_2) 
\left( \frac {8 \as} {\pi} \right) ^2 \CF \\ 
 & &
\frac {x_{10}^2} {x_{30}^2 x_{31}^2}
\left[ \frac {\Nc} {2} 
\left( \frac {x_{30}^2} {x_{43}^2 x_{40}^2} 
     + \frac {x_{31}^2} {x_{43}^2 x_{41}^2} \right) 
     - \frac {1} {2 \Nc} \frac {x_{01}^2} {x_{40}^2 x_{41}^2} \right],
\nonumber
\ea
which in the large $\Nc$ limit reduces to (\ref{onium Phi2}).

\section{Summary}

In this short note we have calculated two soft gluon contribution to
the wave function of the proton. The first gluon emission, described
by eq.(\ref{proton Phi1}), reveals all nice features of the {\em onium}
case: 

1. it is IR finite and 

2. can be interpreted as dipole emission from the initial proton. \\
One has to stress that this result can be obtained only if the interference
diagrams of the type of fig.4b are included. With respect to the graphs of 
the type of fig.4a their color factor is suppressed as $1/\Nc$, however, their
number grows like $\Nc$ and therefore they cannot be neglected.

The emission of the next gluon causes the problem. Kinematical factors
in transverse configuration space corresponding to diagrams of fig.4a 
and fig.4b are completely different. In fact the naive interpretation, that
in the first emission the proton splits into a dipole and another proton
breaks down.  Instead we encounter new color structures, like the one
in fig.4b, where the quark and antiquark lines which are 
localized in the same transverse position are interchanged many times.
This is perhaps the signature that some new color structures, not only
dipoles and three-quark singlets (like proton) appear. However, we do
not have any solid proof of this statement at the moment.

We have also checked by explicit calculation that if 2 (or $\Nc-1$)
quarks are ``by force'' put into the same transverse position the
dipole equation emerges. This is an expected feature, since two 
(or $\Nc-1$) quarks in antisymmetric state behave like an antiquark.

\newpage

\noindent{\Large{\bf Acknowledgements}}

\vspace{0.5cm}

We would like to thank A.~Bia{\l}as for suggesting this investigation
and useful discussions and R.~Peschanski for discussions. The authors
acknowledge the support of Polish KBN Grant PB 2 PO3B 044 12. 
M.P. acknowledges support of A.~v.~Humboldt Foundation.

\end{document}